%% file: 2024asilomar_neuro.tex
\documentclass[conference]{IEEEtran}
\IEEEoverridecommandlockouts
\usepackage{cite}
\usepackage{amsmath,amssymb,amsfonts}
\usepackage{mathtools}
\usepackage{algorithm}
\usepackage{algpseudocode}
\usepackage{graphicx}
\usepackage{textcomp}
\usepackage{xcolor}
\usepackage{hyperref}
\usepackage{pgfplots, tikz}
\usepgfplotslibrary{groupplots}
\usepackage{enumitem}
\usepackage{caption, subcaption}
\pgfplotsset{compat=1.12}

\usepackage[left=0.65in,right=0.65in,top=0.75in,bottom=1.05in]{geometry}
\usepackage{amsthm, bm, siunitx} 
\DeclareSIUnit{\belmilliwatt}{Bm}
\DeclareSIUnit{\dBm}{\deci\belmilliwatt}

\def\BibTeX{{\rm B\kern-.05em{\sc i\kern-.025em b}\kern-.08em
    T\kern-.1667em\lower.7ex\hbox{E}\kern-.125emX}}
\begin{document}

\title{Deep-Unrolling Multidimensional Harmonic Retrieval Algorithms on Neuromorphic Hardware
\thanks{The work
of V.C. Andrei, U.J. Mönich, and H. Boche was supported in part by the German Federal Ministry
of Education and Research (BMBF) within the national initiative on
6G Communication Systems through the research hub 6G-life under
Grant 16KISK002, in part by the German Research Foundation
(DFG) through the Gottfried Wilhelm Leibniz Prize Programme under
Grant BO 1734/20-1, Grant BO 1734/35-1, and Grant STA 864/9-1, and in part by the National Initiative for Quantum Communication in the BMBF Quantum Program QuaPhySI under Grant 16KIS1598K.
The work of A.P. Dragu\c{t}oiu, Y. Li, and U. Mönich was supported by the Bavarian State Ministry
of Science and the Arts through the project GAIn—“Next Generation AI Computing”.
The work of G. B\'{e}na, M. Akl and M. Lohrmann was supported in part by EIC Transition under the ”SpiNNode” project (grant number 101112987) and Horizon Europe project "PRIMI" (Grant number 101120727). 
}
}

\author{\IEEEauthorblockN{
Vlad C. Andrei\IEEEauthorrefmark{1},
Alexandru P. Drăguțoiu\IEEEauthorrefmark{1},
Gabriel B\'{e}na\IEEEauthorrefmark{2}\IEEEauthorrefmark{3},
Mahmoud Akl\IEEEauthorrefmark{2},
Yin Li\IEEEauthorrefmark{1},\\
Matthias Lohrmann\IEEEauthorrefmark{2},
Ullrich J. M\"onich\IEEEauthorrefmark{1},
Holger Boche\IEEEauthorrefmark{1}
\IEEEauthorblockA{\IEEEauthorrefmark{1}Chair of Theoretical Information Technology, Technical University of Munich, Munich, Germany\\
\IEEEauthorrefmark{3}Department of Electrical and Electronic Engineering
Imperial College, London, United Kingdom\\
\IEEEauthorrefmark{2}SpiNNcloud Systems GmbH, Dresden, Germany \\
Emails: \IEEEauthorrefmark{1}\{vlad.andrei, 
alex.dragutoiu,
yin.li,
moenich,
boche\}@tum.de,\\  
\IEEEauthorrefmark{2}\{gabriel.bena, mahmoud.akl, matthias.lohrmann\}@spinncloud.com,\\
}  }}
\maketitle

\begin{abstract}
\input{sections/abstract}
\end{abstract}


\section{Introduction}
\input{sections/paper/intro}
\section{Preliminaries}
\input{sections/paper/sys_model}
\section{Proposed Method}
\input{sections/paper/proposed_method}

\section{Results}
\input{sections/paper/results}
\section{Conclusions}
\input{sections/paper/conclusions}
\bibliography{IEEEabrv, references}
\bibliographystyle{IEEEtran}
\end{document}

%% file: sections/abstract.tex
This paper explores the potential of conversion-based neuromorphic algorithms for highly accurate and energy-efficient single-snapshot multidimensional harmonic retrieval (MHR). 
By casting the MHR problem as a sparse recovery problem, we devise the currently proposed, deep-unrolling-based Structured Learned Iterative Shrinkage and Thresholding (S-LISTA) algorithm to solve it efficiently using complex-valued convolutional neural networks with complex-valued activations, which are trained using a supervised regression objective. 
Afterward, a novel method for converting the complex-valued convolutional layers and activations into spiking neural networks (SNNs) is developed. At the heart of this method lies the recently proposed Few Spikes (FS) conversion, which is extended by modifying the neuron model's parameters and internal dynamics to account for the inherent coupling between real and imaginary parts in complex-valued computations. Finally, the converted SNNs are mapped onto the SpiNNaker2 neuromorphic board, and a comparison in terms of estimation accuracy and power efficiency between the original CNNs deployed on an NVIDIA Jetson Xavier and the SNNs is being conducted. The measurement results show that the converted SNNs achieve almost five-fold power efficiency at moderate performance loss compared to the original CNNs.  

%% file: sections/paper/intro.tex
Multidimensional Harmonic Retrieval (MHR) is a ubiquitous problem in signal processing, finding applications in radar, and wireless joint communications and sensing \cite{liu2022integrated}, among others. Due to the growing bandwidths and high mobility scenarios, developing MHR methods that require one or very few samples is desirable. Compressed sensing (CS) has been an attractive solution due to its good reconstruction qualities and performance guarantees \cite{chi_compressive_2015}. On the other hand, such approaches incur high computational complexity with increased resolution and relatively slow convergence, which impact both latency and energy efficiency. To address the former, the paradigm of deep unrolling  \cite{monga2021algorithm} was proposed, a data-driven, deep-learning (DL) approach, where the iterations of CS are unfolded and parameterized by deep neural networks (DNNs), exhibiting good flexibility and performance with a lower computational burden. Often trained and deployed on Graphics Processing Units (GPUs), these approaches suffer from high energy consumption. To mitigate this, neuromorphic computing platforms such as the SpiNNaker2 \cite{gonzalez_spinnaker2_2024} were developed, which are very much suited for Spiking Neural Network (SNN) computations, an inherently energy-efficient approach inspired by the sparsity in brain computations \cite{eshraghian_training_2023}.
Additionally, it has been shown theoretically that neuromorphic computers have far more representational computers than current digital ones since many highly relevant tasks related to signal processing and communications can not be solved on digital platforms \cite{boche_limitations_2023, boche_remote_2024, boche_denial--service_2020}. 
On the other hand, due to the absence of competitive learning algorithms for SNNs, DNN-to-SNN conversion needs to be performed \cite{eshraghian_training_2023, rueckauer_conversion_2017}. Although effective conversion methods exist, these are mostly limited to real-valued image classification tasks and DNNs with ReLU activation functions \cite{rueckauer_conversion_2017, sengupta_going_2019} and do not support inherent complex-valued regression tasks specific to signal processing applications. 
To our knowledge, there are no spike encoding methods and no network architectures addressing the general processing of complex-valued MHR problems, the closest works being \cite{lopez-randulfe_spiking_2021, lopez-randulfe_time-coded_2022}, where the authors develop a temporal coding approach together with a spiking Fourier transform, which is specialized to compute range-Doppler maps from radar datasets.
This paper explores how deep unrolling-based algorithms for MHR can be deployed on neuromorphic hardware, with our contributions being as follows:
\begin{enumerate}
    \item We propose a novel, complex-valued neuron model based on the few spikes (FS) coding in \cite{stockl_optimized_2021}, which is able to approximate arbitrary, complex-valued activations on a bounded domain using a limited number of spikes.
    \item Using the proposed neuron model, we develop a spike encoding and DNN-to-SNN conversion method for CV-CNNs, which we employ to convert the recently proposed S-LISTA \cite{fu2021structured} to solve general MHR tasks.
    \item Finally, we evaluate the proposed method in terms of task performance and power efficiency on an NVIDIA Jetson Xavier GPU and on the SpiNNaker2 single-chip neuromorphic board \cite{gonzalez_spinnaker2_2024}. Our experimental results show an almost five-fold increase in power efficiency at a moderate task performance loss.
\end{enumerate}

%% file: sections/paper/sys_model.tex
\subsection{System Model and Sparse Recovery}
In the following, we derive a general mathematical formulation of the single-measurement MHR problem as in \cite{dimitri2010tensor} and explain how the framework of sparse recovery can be used to solve it.
Let $\bm{\xi}_{p} = [\xi_{1p}, \xi_{2p}, \dots, \xi_{Mp}]^{\top} \in [-\frac{1}{2}, \frac{1}{2}]^{M}$ be the parameters associated to a source $p=1,\dots, P$. Let $N_{1}, N_{2}, \dots N_{M}$ represent the number of sensors used to capture each dimension and $N = \prod_{m=1}^{M} N_{m}$. 
With $\bm{a}_{m}(\xi_{mp}) \in \mathbb{C}^{N_{m}}$, where
$
    \left(\bm{a}_{m}(\xi_{mp})\right)_{k} = \mathrm{e}^{-\mathrm{j}2\pi \xi_{mp} k}.
$ for $k = 0, \dots N_{m} - 1$,
and the matrices
$\bm{A}_{m}(\bm{\xi}) = [\bm{a}_{m}(\xi_{m1}), \dots, \bm{a}_{m}(\xi_{mP})] \in \mathbb{C}^{N_{m}\times P},
$
the measurement model is given by \cite{dimitri2010tensor}
\begin{align}\label{eq:sys_kr}
    \bm{y} &= \left(\bm{A}_{1}(\bm{\xi}) \ast \cdots \ast \bm{A}_{M}(\bm{\xi})\right)\bm{b} + \bm{z} = \bm{A}(\bm{\xi})\bm{b} + \bm{z}.
\end{align}
Here $\ast$ denotes the Khatri-Rao (KR) product, $\bm{b} = [b_{1}, \dots, b_{P}]^{\top}\in\mathbb{C}^{P}$ are the complex amplitudes and $\bm{z} \sim \mathcal{N}(\bm{0}, \sigma^{2}\bm{\mathrm{I}})\in\mathbb{C}^{N}$. The goal of SS-MHR is to estimate the harmonics $\{\bm{\xi}_{p}\}_{p=1}^{P}$ given the measurement $\bm{y}$. 
Since the number of sources is in general unknown, in the sparse recovery paradigm, $L\gg N$ harmonics $\bm{\xi}_{l}$ are chosen, then an oversampled dictionary $\bm{A}(\bm{\xi}) \in \mathbb{C}^{N\times L}$ is constructed by replacing the KR with the Kronecker-product in Eq. \eqref{eq:sys_kr}. Subsequently, the parameters are estimated from the nonzero indices of the sparse solution found by solving 
\begin{align}\label{eq:p1}
    \min_{\bm{b}\in\mathbb{C}^{L}} \left\lVert \bm{y} - \bm{A}(\bm{\xi})\bm{b} \right\rVert_{2}^{2} - \lambda\lVert \bm{b} \rVert_{1},
\end{align}
where $\lambda > 0$.
This procedure, also known as $\ell_1$-relaxation, theoretically yields near-optimal estimates \cite{beck2009fista}.

A popular approach to solving the problem in Eq. \eqref{eq:p1} is the Fast Iterative and Shrinkage Algorithm (FISTA) \cite{beck2009fista}, which at each iteration $t=1,\dots,T$ constructs a solution as
\begin{align}\label{eq:fista1}
    \bm{b}_t &= \mathcal{S}_{\lambda/\eta}\left(\bm{W}^{1}_{t} \bm{y} + \bm{W}^{2}_{t} \bm{b}_{t-1}\right)\\
    \bm{W}^{1}_{t}&=\eta^{-1}\bm{A}^{H}(\bm{\xi}), \,\bm{W}^{2}_{t} = \bm{I} - \eta^{-1}\bm{A}^{H}(\bm{\xi})\bm{A}(\bm{\xi}) \label{eq:fista2}
\end{align}
where $\eta$ is the largest eigenvalue of $\bm{A}^{H}(\bm{\xi})\bm{A}(\bm{\xi})$ and
\begin{align}\label{eq:st}
    \mathcal{S}_{\alpha}(\bm{x}) = \text{sign}(\bm{u}) \max(\lvert\bm{u}\rvert - \alpha, \bm{0})
\end{align}
is the elementwise soft-thresholding operator with threshold $\alpha > 0$. 
This approach can be seen as a modification of gradient descent, where the solution after each gradient step is projected onto a union of subspaces of restricted support, i.e., with a finite number of non-zero elements.
Even if this approach is theoretically optimal and drastically improves the convergence of traditional ISTA \cite{beck2009fista}, it still needs a very high number of iterations to reach a sparse solution $\hat{\bm{b}}$\cite{fu2021structured}.

\subsection{Structured LISTA}\label{sec:slista}
\begin{figure}
    \centering
    \includegraphics[width=\columnwidth]{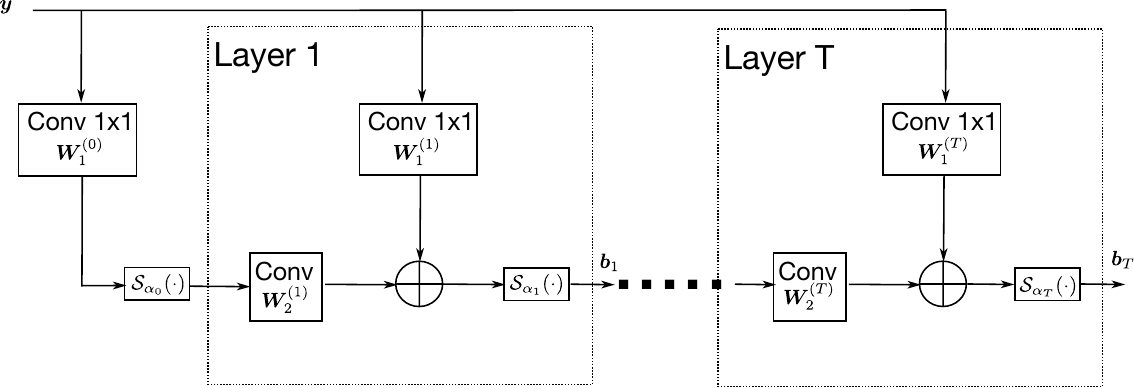} 
    \caption{Schematic representation of S-LISTA architecture.}
    \label{fig:slista}
\end{figure}

In order to address the issue of a large number of iterations in FISTA, the Structured Learned Iterative Shrinkage and Thresholding Algorithm (S-LISTA) was proposed in \cite{fu2021structured}. The algorithm, depicted in Figure \ref{fig:slista} unfolds FISTA recursion in Eqs. \eqref{eq:fista1} to \eqref{eq:fista2} and only considers the first $T$ iterations of the algorithm. It furthermore replaces the matrices $\bm{W}^{1}_{t},\bm{W}^{2}_{t}$ with trainable CNNs and considers the threshold $\alpha_t = \lambda/\eta$ also as a trainable parameter. More formally,
\begin{align}\label{eq:slista}
    \bm{b}_t &= \mathcal{S}_{\alpha_t}\left(
    \text{Conv1x1}_{\bm{W}^{1}_{t}} (\bm{y}) + \text{Conv}_{\bm{W}^{2}_{t}}(\bm{b}_{t-1})\right)
\end{align}
where $\text{Conv1x1}_{\bm{W}^{1}_{t}}$ is 1x1-convolution \cite{chollet_xception_2017} (equivalent to matrix-vector multiplication) and $\text{Conv}_{\bm{W}^{2}_{t}}$ 
is the convolution operation. 
The proposed architecture is motivated by the fact that the matrix $\bm{W}^{2}_{t}$ is block-Toeplitz, which allows the matrix-vector multiplication in Eq. \eqref{eq:fista2} to be implemented efficiently with $M$-D convolutions with separable kernels, 
which is a fast operation on modern GPUs. We refer the reader to \cite{fu2021structured,chi_compressive_2015} for detailed derivations.
At last, the well-known equivalence between matrix-vector multiplications and 1x1-convolutions \cite{chollet_xception_2017} not only theoretically underpins the layer computations in Eq. \eqref{eq:slista}, but also allows for fast and efficient training on GPUs.

In order to train the network, we first generate a labeled training dataset $\mathcal{D} = \{(\bm{y}_{d}, \bm{b}_{d})\}_{d}^{D}$ for multiple numbers of sources $P$ and noise variances $\sigma^2$. We do this by using a normalized version of Eq. \eqref{eq:sys_kr} to obtain a sample-wise signal-to-noise ratio (SNR) equal to $1/\sigma^2$.
We train the network by back-propagation using the ADAM optimizer \cite{kingma_adam_2017} with the following modified objective:
\begin{align}
    \mathcal{L}(\bm{b}_d , \hat{\bm{b}}_d) = \frac{1}{D}\sum_{d=1}^{D} \dfrac{\lVert\bm{b}_d -  \hat{\bm{b}}_d \rVert_{1}}{\lVert\bm{b}_d \rVert_1}.
\end{align}
where $\hat{\bm{b}}_d$ denotes the network output.
In \cite{fu2021structured}, the $\ell_2$-norm was used, but we empirically observed that the $\ell_1$-norm helps us achieve faster training convergence and better enforces sparse structure.
\subsection{Spiking Neural Networks and SpiNNaker2}
In order to explain the conceptual difference between artificial neural networks (ANNs) and SNNs, we recall the fact that ANNs process a real(or complex)-valued input vector $\bm{x}\in\mathbb{R}^{D}$ in using the iterated composition of element-wise nonlinear functions, called activations, with affine-linear maps \cite{boche_limitations_2023}. 

In contrast, SNNs do not operate on real-valued inputs but rather on a time series of spikes, which are then processed by following the dynamics of a neuron model instead of being passed to an activation function. More formally, for the well-known leaky-integrate-and-fire (LIF) neuron, the internal dynamics for $K$ timesteps are given by \cite{eshraghian_training_2023}:
\begin{align}
    \bm{v}_n(t) &= \beta\bm{v}_n(t-1) + \bm{W}_n \bm{z}_{n-1}(t) + \bm{z}_{n}(t-1) V_{th}\\
    \bm{z}_{n}(t) &= \Theta(\bm{v}_n(t) - V_{th}), \quad t=0,\dots, K-1
\end{align}
where $\bm{z}_{n}(t) \in \{0,1\}^{D_n}$, $\bm{v}_n(t)\in\mathbb{R}^{D_n}$, $\bm{W}_n \in \mathbb{R}^{D_n \times D_{n-1}}$ denote the spikes, membrane potentials and weights at time $t$ for the $n$-th layer, with the element-wise Heaviside function, i.e. $\Theta(x) = 1, x>0$ and $0$ otherwise. Here $\beta>0, V_{th}\geq0$ is the leakage factor and the threshold potential, which determines whether the neuron fires or not. 

In order to process real- or complex-valued data with SNNs, one first needs to encode the input $\bm{x}$ into an initial spike train $\bm{z}_0(t)$, with rate- and latency-encoding being the most popular methods \cite{eshraghian_training_2023}.   
Furthermore, note that the firing rate of the LIF neuron with rate-coded spikes
is described by the rectified linear unit (ReLU) \cite{rueckauer_conversion_2017}. 
Thus, in terms of converting an ANN to an SNN, the representational power of the LIF neuron is restricted to real-valued SNNs employing ReLU activation functions \cite{rueckauer_conversion_2017, sengupta_going_2019}.

SpiNNaker2 is a massively parallel, highly scalable and energy-efficient neuromorphic system that takes a hybrid approach to neuromorphic computing by allowing both spiking and classical rate-based neural networks to be executed onto the board \cite{gonzalez_spinnaker2_2024}.
One chip contains 152 ARM Cortex-M4 cores, also called Processing Elements (PEs),
which can host at least 250 neurons with a total number of 1000 synapses per neuron. Each PE runs a small asynchronous program and communicates with other PEs via Network-on-Chip (NoC) components, which also facilitates process scheduling and dynamic voltage and frequency scaling (DVFS) in order to allocate resources only if an operation occurs.

%% file: sections/paper/proposed_method.tex
\subsection{Complex-Valued FS Coding}

In order to leverage the enhanced parallelism and energy efficiency of neuromorphic platforms, we convert the trained models to SNNs. For this, the complex values and activations need to be converted to a spiking domain following a spike coding scheme. We propose an extension of the FS coding scheme \cite{stockl_optimized_2021}, which utilizes two distinct FS neurons, one for encoding the real part and another for the imaginary part. Each of these neurons transmits serialized binary codes over a $K$ timesteps, which aggregated approximate an arbitrary activation $f(s) : \mathcal{B} \mapsto \mathbb{C}$ over a bounded domain $\mathcal{B} \subset \mathbb{C}$. The internal dynamics of a single neuron are governed by:
\begin{align}\label{eq:fs_cmplx}
    v_{\Re}(t + 1) &= v_{\Re}(t) - h_{\Re}(t)z_{\Re}(t), \\ v_{\Im}(t + 1) &= v_{\Im}(t) - h_{\Im}(t)z_{\Im}(t),\\
    \label{eq:fs_cmplx2}
    z_{\Re}(t) &= \Theta(v_{\Re}(t) - T_{\Re}(t)),\\ z_{\Im}(t) &= \Theta(v_{\Im}(t) - T_{\Im}(t)).
\end{align}
In this formulation, $d_{\Re}, h_{\Re}, T_{\Re}, d_{\Im}, h_{\Im}, T_{\Im} \in \mathbb{R}^K$ represent parameters associated with the real and imaginary parts of the FS neurons.
Furthermore, $v_{\Re, \Im}(t)$ and $z_{\Re, \Im}(t)$ represent the membrane potentials and the generated spikes of two FS neurons. Note that the original formulation for real-valued processing in \cite{stockl_optimized_2021} is recovered by only considering Eqs. \eqref{eq:fs_cmplx} and \eqref{eq:fs_cmplx2}.
The output of the neuron, which approximates the original activation, is then computed as:
\begin{align}
    \hat{f}(s) = \sum_{t=0}^{K-1} (d_{\Re}(t) z_{\Re}(t) + \mathrm{j} \cdot d_{\Im}(t) z_{\Im}(t)).
\end{align}
In order to get a good approximation over $\mathcal{B}$, $B$ points are sampled, and the following optimization problem is solved by gradient descent and a triangular approximation for the subgradient of $\Theta(\cdot)$ as done in \cite{stockl_optimized_2021}
\begin{align}\label{eq:fs_optimization}
    \min_{\substack{d_{\Re}, h_{\Re}, T_{\Re},\\ 
    d_{\Im}, h_{\Im}, T_{\Im} \in \mathbb{R}^K}} 
    \sum_{i=1}^{B} \lvert f(s_{i}) - \hat{f}(s_{i}) \rvert^{2}.  
\end{align}

The spike processing mechanism of FS neurons is characterized by two cyclic stages. Specifically, these neurons undergo a stage of spike accumulation spanning $K$ timesteps followed by an additional stage of spike transmission spanning another $K$ timesteps. This behavior effectively enables each neuron to process new input at regular intervals of $2K$ timesteps, as the neurons remain idle after each receiving-transmitting cycle. Consequently, this operational paradigm improves throughput capacity within the neural network architecture.

\subsection{ANN-to-SNN Conversion Pipeline}
In the following, we describe how the S-LISTA network presented in Section \ref{sec:slista} is converted to an equivalent SNN based on complex FS coding. In principle, each layer implements the same functional form as \eqref{eq:slista} but with the activations replaced by their FS approximations. Furthermore, the spike encoding is done by passing the input data to the FS approximation of the identity. More formally,
\begin{align}
    \bm{b}_t &= \text{FS}_{\mathcal{S}_{\alpha_t}}\left(
    \text{Conv1x1}_{\bm{W}^{1}_{t}} (\text{FS}_{\text{Id}}(\bm{y})) + \text{Conv}_{\bm{W}^{2}_{t}}(\bm{b}_{t-1})\right),
\end{align}
where $\text{FS}_{\mathcal{S}_{\alpha_t}}$ and $\text{FS}_{\text{Id}}$ are obtained by solving \eqref{eq:fs_optimization} for the function in Eq. \eqref{eq:st} and $f(s) = s$, respectively.


\begin{algorithm}[t]
\caption{S-LISTA Conversion and Mapping SpiNNaker2}
\label{alg:conversion}
\textbf{Input:} Weights $\bm{W}^{1}_{t}, \bm{W}^{2}_{t}$, thresholds $\alpha_t$.
\vspace{1mm} \hrule \vspace{1mm}
\begin{algorithmic}[1]
\item \label{alg:step0} Solve \eqref{eq:fs_optimization} to obtain $\text{FS}_{\text{Id}}$.
\For{$t = 1, \dots ,T$}
\item \label{alg:step1} Solve \eqref{eq:fs_optimization} to obtain $\text{FS}_{\mathcal{S}_{\alpha_t}}$.
\item Perform 4-bit-quantization of $\bm{W}^{1}_{t}, \bm{W}^{2}_{t}$ and create projections for convolution operations for real and imaginary parts separately.
\item \label{alg:step3} Create populations and their delay for $\bm{b}_{t}$ and $\bm{y}$. 
\item Send the objects in Steps \ref{alg:step1} and \ref{alg:step3} to SpiNNaker2's PEs.
\EndFor
\item At inference, encode inputs into spikes with $\text{FS}_{\text{Id}}$ and run programs on SpiNNaker2's PEs.
\end{algorithmic}
\end{algorithm}
In order to understand the steps of the full ANN-to-SNN conversion in Algorithm \ref{alg:conversion}, it is important to understand how FS neurons are implemented on SpiNNaker2. On the board, FS neurons within a population, i.e. internal data structure representing the spiking neurons for $\bm{y}$ and $\bm{b}_{t}$, generate and process spikes in parallel. Once the simulation begins, the populations enter a synchronized receive-send cycle. This ensures that spikes do not travel across the network continuously, as the receiving population is inactive while the transmitting population is. To address this, spike transmission in the previous layer is delayed by a stage duration $K$ to synchronize spike arrival with the receiving stage of the next layer. This approach enables pipelining, allowing neurons to process new input after $2K$ timesteps. During inference, input data is encoded into spikes using trained FS codes. An offset of $2K t$ is added to each spike, where $t=0,\dots, T-1$, ensuring timely arrival for each layer’s FS populations.

Finally, we note that due to both hardware and software constraints, the projections, i.e. the data structure representing the connectivity information between populations extracted from the weights, are constrained to $4$ bits. 
To run the SNNs on SpiNNaker2, it is thus necessary to quantize the real and imaginary parts of $\bm{W}^{1}_{t}, \bm{W}^{2}_{t}$, which we do after training. Note that this is only necessary for our models to run on the current experimental version of the single-chip board and does not constitute a limitation of our approach.

%% file: sections/paper/results.tex
\begin{figure}
    \centering
    \begin{subfigure}{\columnwidth}
        \centering
        \includegraphics[width=\columnwidth]{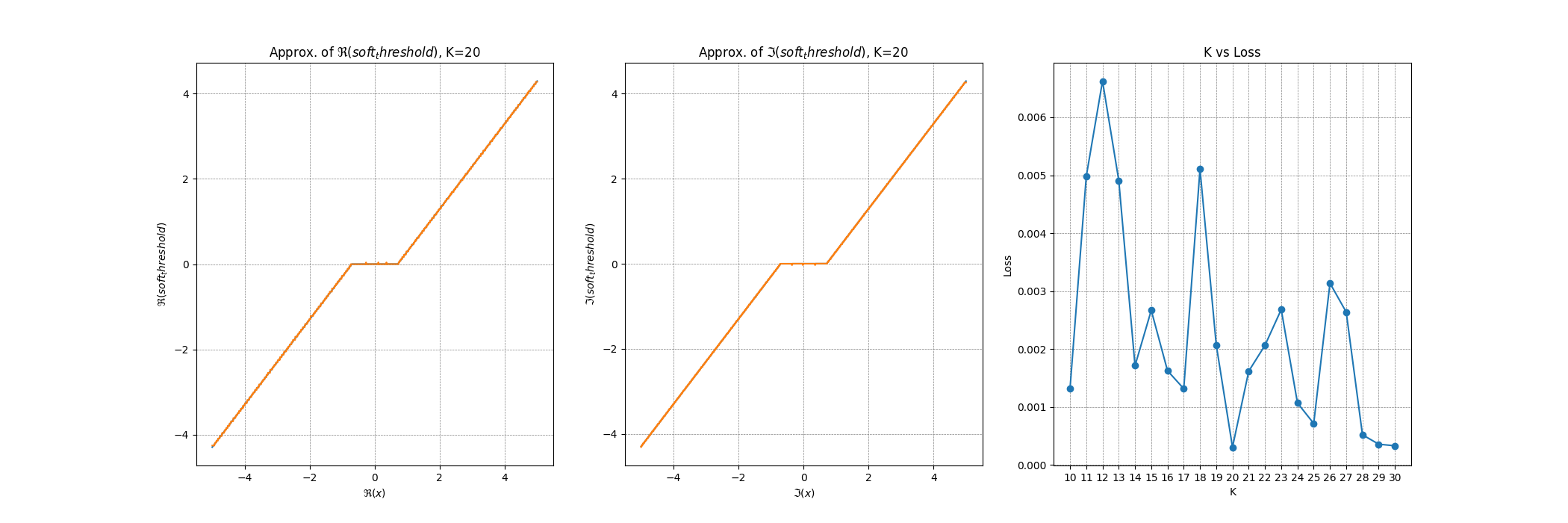}
    \end{subfigure}
    \begin{subfigure}{\columnwidth}
        \centering
        \includegraphics[width=\columnwidth]{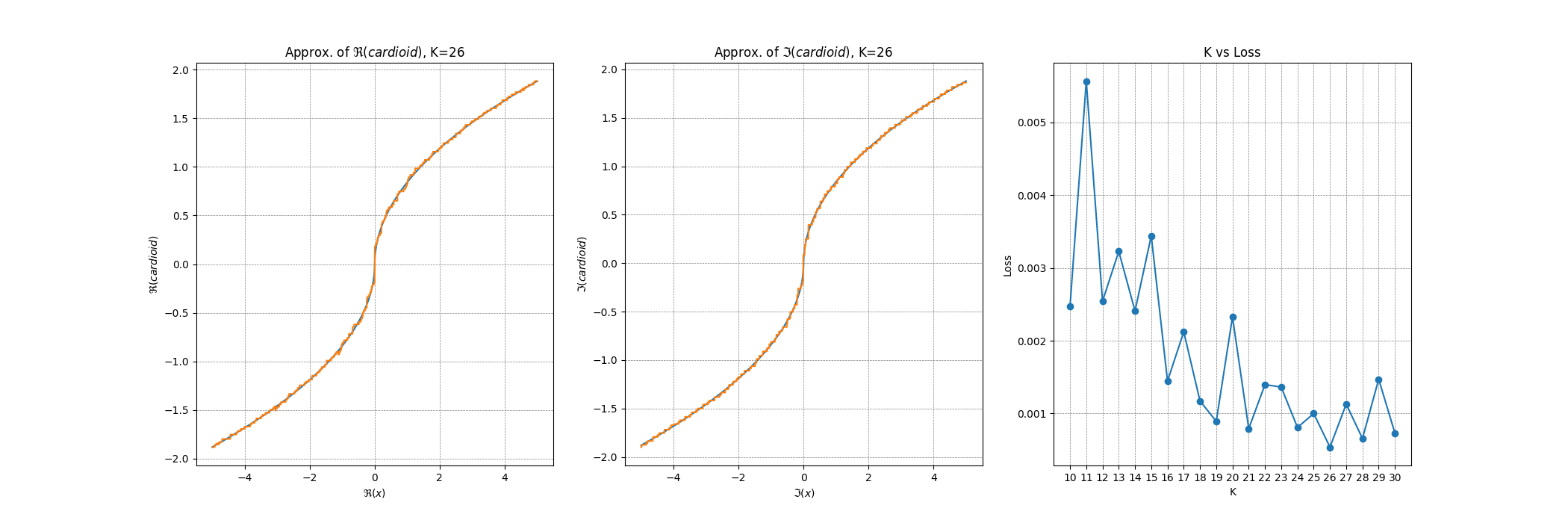}
    \end{subfigure}
    \caption{FS neuron approximating Eq. \eqref{eq:st} ($\alpha = 1$) and $f(s)=s^{2}$ for different $K$. Blue denotes the true function, and orange is the approximation.}
    \label{fig:fs_training}
\end{figure}

We first evaluate the proposed neuron model in terms of function approximation capabilities. Figure \ref{fig:fs_training} shows the real and imaginary parts of the true and approximated function, as well as the error for different timesteps $K$.
For all trials, a learning rate of $\beta = 0,009$ and $I=10000$ iterations were employed for the gradient descent solver.    
We observe that the proposed method qualitatively and quantitatively approximates the functions quite well. The real and imaginary parts look very similar, and the value of the optimization criterion in Eq. \ref{eq:fs_optimization} stays below $8\cdot 10^{-3}$. Furthermore, the approximation error tends to decrease with the number of timesteps $K$. We thus observe that the required $K$ to get a good approximation is below $30$, which makes the method promising regarding delay and energy efficiency, since it is much less than the hundreds of timesteps LIF neurons usually employ  \cite{rueckauer_conversion_2017}.

\begin{figure}
    \centering
    \includegraphics[width=.7\columnwidth]{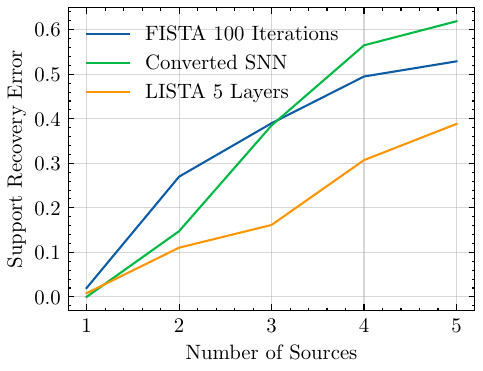}
    \caption{SRE averaged over SNRs from $10\text{dB}$ to $30\text{dB}$.}
    \label{fig:performance}
\end{figure}

We now present the task performance results for the converted SNNs on the SpiNNaker2 neuromorphic board. Due to the very strict memory limitations of the 1-chip evaluation board, we were only able to map S-LISTA models for $M=1$-D HR problems with a maximum of $T = 5$ layers and an observation size of $N = 64$ and dictionary size $L = 128$. We have trained the S-LISTA networks for $30$ epochs using a learning rate of $5\cdot10^{-4}$. For the ANN-to-SNN conversion pipeline, we have used a learning rate of $10^{-2}$ for the optimization in \eqref{eq:fs_optimization} and terminated the SGD solver once an error below $5\cdot10^{-4}$ was reached.
Figure \ref{fig:performance} shows the support recovery error of the FISTA algorithm, the S-LISTA networks, and the converted SNN for $P=1$ to $5$ sources, which is given by 
\begin{align}
 \text{SRE} =
  \frac{1}{D}\sum_{d=1}^{D} \frac{\left|\{l \neq l'\ | l \in \text{supp}(\bm{\hat{b}}_d), l' \in \text{supp}(\bm{b}_d)\}\right|}{
\text{supp}(\bm{b})
},
\end{align}
where $\text{supp}(\bm{b}) = \{l\in\mathbb{N}\, |\, \lvert\bm{b}_l \rvert \geq 0\}$. We observe that the converted SNN outperforms FISTA and performs similarly to S-LISTA for a few sources. The SNN performance worsens with an increasing number of sources but still shows comparable results to the original FISTA formulation at a fraction of the resources, i.e., $5$ layers with less complexity compared against the $100$ iterations needed for FISTA. 
This is mainly explained by quantization effects and the residual errors in the FS neuron approximations. We thus observe that quantization is the bottleneck to good performance and that the employed post-training quantization is not optimal. Furthermore, the FS approximation artifacts can be partly mitigated by employing a bigger $\beta$ and more finely sampling $\mathcal{B}$.
We now turn our attention to the power efficiency measurements on the NVIDIA Jetson Xavier NX \cite{nvidia2024}, and the experimental SpiNNaker2 \cite{gonzalez_spinnaker2_2024} boards. 
The power efficiency evaluation procedure is done as follows: The host PC streams data to both devices for inference in one thread, while in the other thread, the values of the instantaneous powers are recorded and stored. For the NVIDIA Jetson Xavier, we measure the consumed power for S-LISTA from the common \texttt{CPU/GPU} rail \cite{nvidia2024} using the \texttt{tegrastats}-API. For SpiNNaker2, we obtain the relevant measurements for the converted SNN from the \texttt{V05/V08} \cite{gonzalez_spinnaker2_2024} power rails using software provided by SpiNNcloud GmbH.  
In order to get reliable results, we isolated inference events on both boards from all other processes by incorporating sufficient sleep times, as well as estimated and subtracted the power consumption in idle mode.

\begin{figure}
    \centering
    \includegraphics[width=.7\columnwidth]{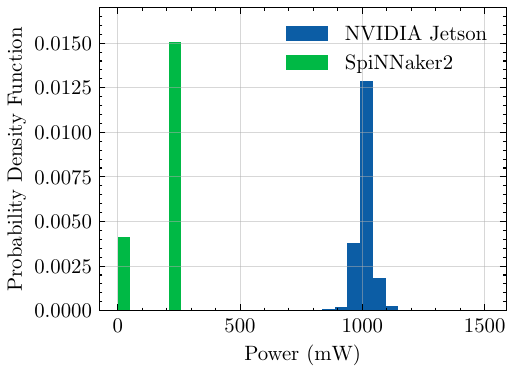}
    \caption{Power consumption statistical distribution.}
    \label{fig:power_eff}
\end{figure}

Figure \ref{fig:power_eff} shows the probability density function (PDF) of the consumed powers on the aforementioned devices.
For NVIDIA Jetson we see a PDF centered around $\SI{1022,35}{\milli\watt}$ with a standard deviation of $\SI{35,95}{\milli\watt}$, while for the SpiNNaker2 the PDF consist of two spikes, one centered around $\SI{25}{\milli\watt}$, the other one around $\SI{225}{\milli\watt}$. 
The two spikes in the distribution are explained by the DVFS method used in the chip, which only allocates clock frequency and voltage resources to the processing elements if an operation occurs.
By comparing the empirical means, we observe that, on average, the SpiNNaker2 board consumes almost $\SI{768}{\milli\watt}$ less power than the embedded GPU, showing an almost $5\times$ increase in power efficiency.
Preliminary inference delay measurements also show that inference takes roughly the same time on both devices, which leads us to believe that the neuromorphic board attains much better energy efficiency at a moderate performance loss.


%% file: sections/paper/conclusions.tex
We casted the MHR problem as a sparse recovery problem, and solved it using the recently proposed S-LISTA algorithm. We developed a novel method for converting the complex-valued convolutional layers and activations in S-LISTA into SNNs based on the recent FS conversion.
At last, the converted SNNs were mapped onto the SpiNNaker2 neuromorphic board, and a comparison in terms of support recovery error and power efficiency between the original CNNs, deployed on embedded GPU, and the SNNs was conducted. The measurement results show that the converted SNNs can achieve much better power efficiency at moderate performance loss compared to the original CNNs.  